\documentclass[fleqn,twoside]{article}
\usepackage{espcrc2,epsfig}

\usepackage{graphicx}
\usepackage[figuresright]{rotating}

\newcommand{\AmS}{{\protect\the\textfont2
  A\kern-.1667em\lower.5ex\hbox{M}\kern-.125emS}}
 
\title{
Search for the Lepton Flavor Violating B Decays $B^0 \to \mu \tau$ and $B^0 \to e\tau$ at CLEO2
}

\author{J.E. Duboscq\address{ Wilson Laboratory, Cornell University, Ithaca NY 14850, USA}
        \thanks{I am thankful for the fine efforts of the CESR staff for
         creating the many collisions which make this work possible. I would also like to
          thank the National Science Foundation for its essential support.
          Presented at the 8th International Workshop on Tau-Lepton Physics, Nara, Japan, 
         Sept 2004. To appear in Nuclear Physics B - Proceedings Supplements.
}
          for the CLEO collaboration.
          }

\begin{document}

\begin{abstract}
Using almost 10 million $B\overline{B}$ decays produced at CESR, the CLEO collaboration has
 set the most stringent limits to date on the lepton flavor violating decays $B^0 \to \mu (e) \tau$.
\vspace{1pc}
 \end{abstract}

\maketitle

\section{Introduction}
A lepton flavor violating decay such as $ B^0 \to \mu \tau$ and $B^0 \to e \tau$ could conceivably proceed through
 a box diagram involving an exchange of a neutrino along with a suitable mixing matrix. In the 
 Standard Model, neutrinos are usually taken to be massless, and this 
   decay route is thus excluded. One expects that with neutrino masses in the eV range, this
   decay would be heavily suppressed and should not visible at any current facilities. Thus observation
    of this decay mode would constitute important evidence of physics beyond the Standard Model.
    The previous best limits~\cite{oldcleo2result} on the branching fractions are $B(B^0 \to \mu \tau ) < 8.3 \times 10^{-4}$ and
     $B(B^0 \to e \tau) < 5.3 \times 10^{-4}$.

    The data used in this analysis were produced at the CESR $e^+e^-$ collider at the $\Upsilon (4S) $ resonance 
    using the CLEO2 detector~\cite{cleodet}. 
    The dataset comprises some $9.6\times 10^6 B\overline{B}$ decays, and also includes some $4.5 fb^{-1}$ of data 
    taken some $60 MeV$ below the resonance
     to gauge the size of non B backgrounds, including continuum $q\overline{q}$ production, and
      2 photon fusion events. The continuum (and tau) event samples below resonance can be scaled according to the 
      ratio of accepted luminosity to beam energy squared, ${\cal L} /E_{Beam}^2$, to extrapolate 
      the behavior of these components at the $\Upsilon(4S)$. The scale factor is 1.99 for this data.
      Backgrounds from 2 photon events are minimized by doing a cut on the missing momentum
     in the event.
       
       The final state events we look for involve $\mu(e)^+\tau^-$ or its charge conjugate recoiling against a generic $B^0$ or $\overline{B}^0$ 
        decay. 
       The $\tau$ is identified by its leptonic decays to $e\nu\nu$ and $\mu\nu\nu$ states. As a shorthand for this presentation, 
        we refer to the 
       decay chain $B\to l \tau, \tau\to l' \nu \nu$ as $(l,l')$. In the B rest frame, the primary lepton, $l$, is monoenergetic. In the lab
        frame, the $B$ is slightly boosted, and thus the primary lepton has a momentum between 2.2 and 2.5 GeV, easily within the
        electron and muon particle identification abilities of the CLEO detector. The secondary lepton has a range of momenta and is 
        required to have momentum greater than 0.6 (1.0) GeV for electron (muon) identification.
        In this study the missing energy-momentum four vector, determined by reference to the known beam energy, is referred 
        to as $P_{\nu\nu}^\mu$.  
        
         Two neural networks are used to exclude backgrounds from continuum and $B\overline{B}$ events.
         For continuum suppresion, the neural net $NN_{cont}$ uses as inputs
         R2 (the ratio of the 2nd to 0th Fox Wolfram moments), the event sphericity, the event thrust,
         the cosine of the angle between the momentum difference of the leptons and the thrust axis of the rest of the event, 
         and the cosine of the angle between the (unobserved) neutrino pair and lepton pair. This network is trained using samples of 
         signal and generic continuum Monte Carlo. 
         The network used to suppress events from $B\overline{B}$ events, $NN_{BB}$, uses three inputs.
          These are the beam energy constrained B candidate mass, the difference between the  candidate B energy and the beam energy, 
          and the cosine of the 
          angle between the primary lepton and the momentum of the other $B$ in the event.
          This network is trained with signal and $B\overline{B}$ generic Monte Carlo samples.
          For each of the four possible $(l,l')$ modes, events are rejected by cuts in the $NN_{cont}$ vs $NN_{BB}$ plane. 
 
\begin{table*}[htd]
\caption{ Summary of the data for each mode: Numbers of observed events on resonance (N(on)), and off resonance (N(off)) for the primary lepton signal region, number of events 
 remaining after on resonance subtraction of off resonance events in the primary lepton signal region (N(obs)), and the expected numbers  of events from ${B\overline{B}}$ and continuum Monte Carlos. The final 90 $\%$ confidence level upper limits for each mode are also
   given.
 }
\begin{center}
\begin{tabular}{|c|c|c|c|c|} \hline
           $(l,l')$ & $ (\mu,e)$  & $(\mu,\mu)$ & $(e,e)$ & $(e,\mu)$ \\ \hline 
            N(on) 	& $19$	& $ 10	$&	$28$& $6$ \\  \hline
            N(off)	&  $2$	& $3$	& $ 7$& $0$ \\  \hline
            N(obs) & $15.0 \pm 5.2$ & $4.0 \pm 4.7$ & $14.0\pm7.5$ & $6.0 \pm 2.4$ \\ \hline
            $<N_{BB}>$ & $23.7\pm 2.7 $& $9.0 \pm 1.4$ & $11.6\pm1.4$ & $5.1\pm 0.8$ \\ \hline
            $<N_{cont}>$ & $1.8 \pm 0.6$ & $0.4 \pm 0.2$ & $3.1\pm1.0$ & $0.5 \pm 0.3$ \\ \hline
                    U.L. ($10^{-4}$) 	& $0.55$	& $ 0.87	$&	$1.64$& $1.46$ \\  \hline
\end{tabular}
\end{center}
\label{table:1}
\end{table*}

\begin{figure}[htb]
\psfig{figure=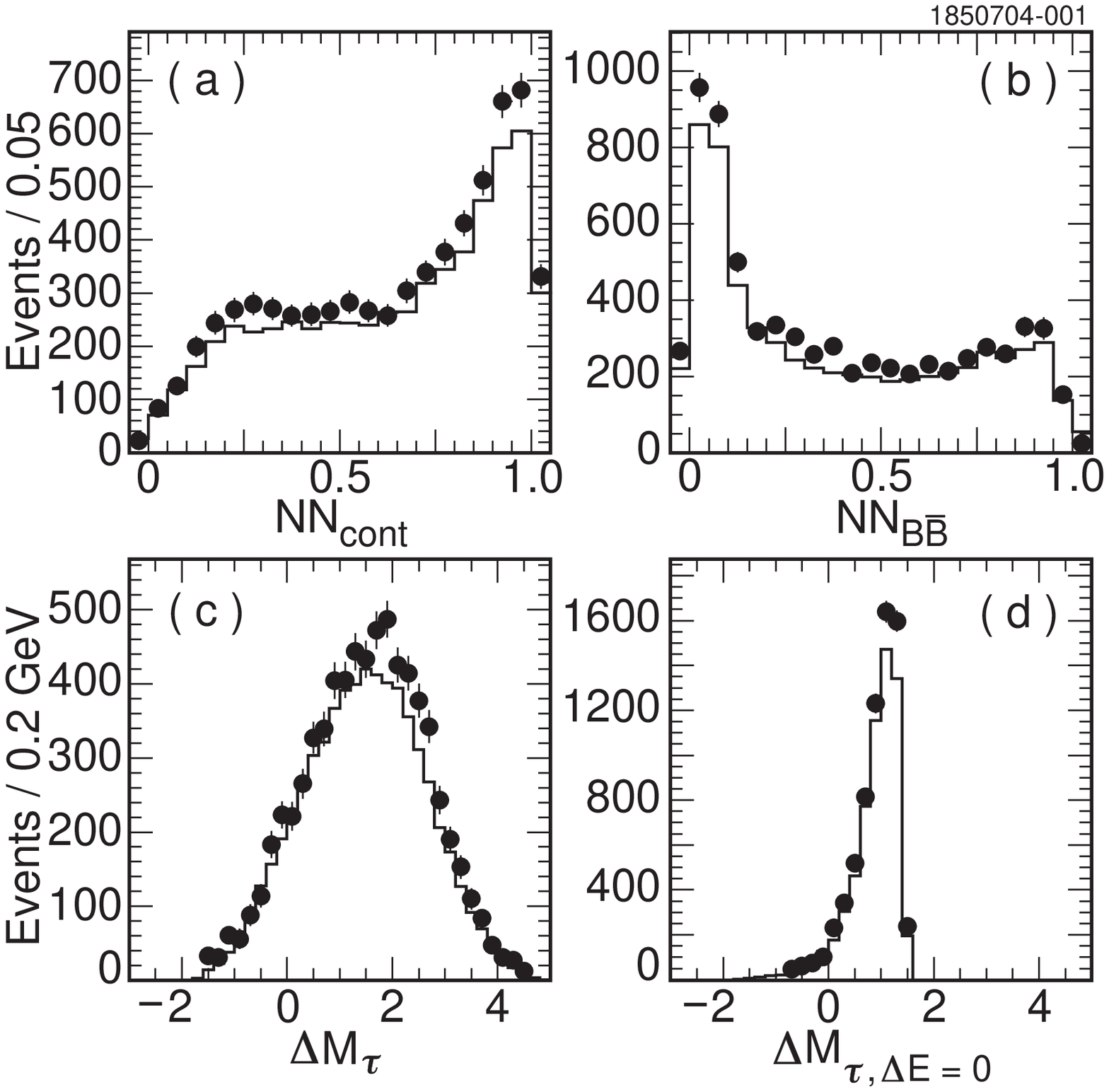, ,scale=0.40}
\vskip -1cm
\caption{ (a) $NN_{cont}$, (b) $NN_{BB}$, (c) measured $\tau$ mass, (d) beam energy constrained $\tau$ mass distributions for ${B\overline{B}}$ Monte Carlo (histogram) and off-resonance subtracted data (points) in the primary lepton sideband region for the $(\mu,e)$ mode.
}
\label{fig:fig1}
\end{figure}

           As a cross check of the analysis method, we compare the data and Monte Carlo agreement in the side bands of the primary 
           lepton - this sample requires that the primary lepton momentum be either in the (2.0,2.2) GeV window or the (2.5,2.7) GeV window.
           Fig~\ref{fig:fig1} is a plot of the off resonance subtracted Data and generic B Monte Carlo for events in this primary lepton sideband for 
           the $(\mu,e)$ sample. The plots show the output of the continuum neural net, the $B\overline{B}$ neural net, the difference between the 
           measured and true $\tau$ mass using $P_{\nu\nu}^\mu$, and the beam energy constrained $\tau$ mass difference, in which
            the missing neutrino pair energy is taken to be the difference between the beam energy and the sum of the lepton energies.
           
\begin{figure}[htd]
\psfig{figure=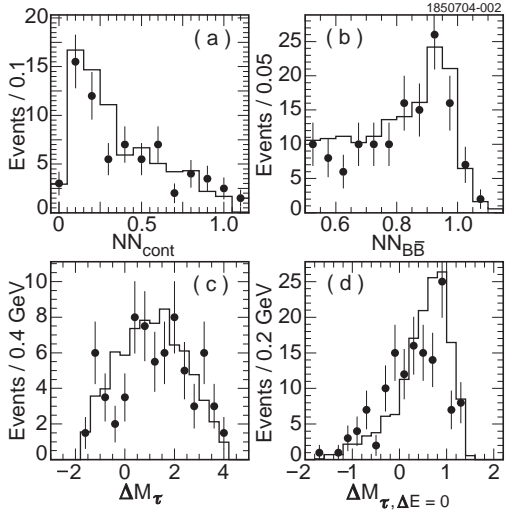,scale=0.40}
\vskip -1cm
\caption{$NN_{cont}$, (b) $NN_{BB}$, (c) measured $\tau$ mass, (d) beam energy constrained $\tau$ mass distributions for continuum Monte Carlo (histograms) and off resonance data (points) in the primary lepton sideband region for the $(\mu,e)$ mode.
}
\label{fig:fig2}
\end{figure}

           Another check  compares primary lepton sideband data for the off resonance data events to the 
           events in the absolutely normalized continuum generic Monte Carlo. Fig~\ref{fig:fig2} shows the same quantities 
           as above for these samples in the $(\mu,e)$ 
           sample. The agreement is good. It is found however that in the $(e,e)$ mode, the data exceeds the continuum 
           Monte Carlo expectation, due to the presence of unmodelled 2 photon fusion events. We thus scale the Monte Carlo 
            in the signal region by this ratio. The contribution to the final error is 
           small since we will end up doing  a subtraction of the scaled off resonance data from
            the on resonance data.

            To obtain the final answer, we subtract the off resonance data from the on resonance data after scaling for center of mass 
            energy evolution with ${\cal L}/E_{Beam}^2$. This is then compared to the expected contribution from $B\overline{B}$ 
             and continuum, which is 
            reckoned by looking at the primary lepton sidebands in both data and Monte Carlo.  Any excess would be signal. 
            The results from individual channels are in Table~\ref{table:1}. The table shows no substantial deviation from expectations.
            
Upper limits derived from the raw data and Monte Carlo efficiencies are also shown in Table~\ref{table:1}. These limits
 include variations on all signal and background efficiencies, and conservatively scale the Monte Carlo in the
  least favorable direction by 1 $\sigma$. The largest single systematic uncertainty is in the estimation of the missing four momentum 
   of the two neutrinos ($5.4\%$), resulting in total systematic uncertainties of $7.4\%$ in $(\mu,e)$ and $(e,\mu)$ modes and
    $8.9\%$ in $(e,e)$ and $(\mu,\mu)$ modes.


Combining these results we obtain the $90\%$ C.L. upper limits on the branching fractions
 to be :
 \begin{eqnarray*}
 B(B^0 \to \mu\tau ) &<& 3.8\times 10^{-5}  \\
  B(B^0 \to e\tau ) &<& 1.1\times 10^{-4}  
  \end{eqnarray*}

These results assume that the $\tau$ from these decays are unpolarized. In the limit of a 
 pure $V-A$ ($V+A$) interaction the efficiency increases (decreases) by $11\%$ ($8\%$).
These are the most stringent limits to date on these processes, surpassing the previous CLEO
 results~\cite{oldcleo2result} by a factor of 22(5) for the $\mu$($e$) primary lepton mode. This work has been submitted
  to Phys. Rev. Lett.

\end{document}